\documentclass[amssymb,amsmath,prb,twocolumn,superscriptaddress,floatfix,showpacs]{revtex4}
\usepackage{graphicx}

\begin{document}
\title{Electrical readout of a spin qubit without double occupancy.}

\author{Andrew D. Greentree}

\affiliation{Centre for Quantum Computer Technology, School of
Physics, The University of New South Wales, Sydney, NSW 2052,
Australia.}
\affiliation{Centre for Quantum Computer Technology, School of
Physics, University of Melbourne, Melbourne, Vic. 3010,
Australia.}

\author{A. R. Hamilton}
\affiliation{Centre for Quantum Computer Technology, School of
Physics, The University of New South Wales, Sydney, NSW 2052,
Australia.}

\author{Lloyd C. L. Hollenberg}
\affiliation{Centre for Quantum Computer Technology, School of
Physics, University of Melbourne, Melbourne, Vic. 3010,
Australia.}

\author{R. G. Clark}
\affiliation{Centre for Quantum Computer Technology, School of
Physics, The University of New South Wales, Sydney, NSW 2052,
Australia.}

\date{\today}

\begin{abstract}
We identify a mechanism to read out a single solid-state electron
spin using an all-electrical spin-to-charge conversion in a closed
system. Our scheme uses three donors and two electron spins, one
spin is the qubit, the other is a reference. The population in the
third, originally ionized, donor is monitored with an electrometer.
Energy dependent tunneling of the reference spin to the ionized
donor is used to determine the state of the qubit. In contrast to
previous methods [e.g. Kane, Nature (London), \textbf{393}, 133
(1998)] we avoid double electron occupancy of any site within the
system, thereby eliminating the possibility of unwanted electron
loss from the system. The single spin readout scheme described here
is applicable to both electron and nuclear spin based quantum
computer architectures.
\end{abstract}


\maketitle


Understanding, observing and manipulating the quantum coherent
properties of individual spins is an important endeavor for the
physics community.  Spin systems offer a superb probe of
fundamental quantum properties.  As such, they have been suggested
and employed in various flavors as elements for quantum computers
(QCs). It is believed by many that one of the best systems for
realizing a scalable and practical spin-based QC is a solid-state
system, that is fully compatible with existing technologies. This
philosophy is encapsulated in several proposals, including the
Kane proposal \cite{bib:Kane}, which uses the nuclear spins of
phosphorus in isotopically pure $^{28}$Si as the qubits; the Loss
and DiVincenzo \cite{bib:LossPRA1998} approach, where the qubits
are electron spins in single-electron quantum dots; and the
electron-spin-resonance approach of Vrijen \textit{et al.}
\cite{bib:VrijenPRA2000}. Progress towards realizing the Kane
device has been recently reviewed \cite{bib:KaneProgress2003}.

Readout of a scalable spin-based QC relies on the ability to sense
the state of single spins. The search for effective methods to
measure single spins has involved researchers from many different
disciplines.  Numerous techniques have been suggested, including
electrical spin-to-charge conversion \cite{bib:Kane,bib:SpinChargeNature2004}, spin
amplification using a paramagnetic dot \cite{bib:LossPRA1998},
spin valves \cite{bib:LossPRA1998}, magnetic-resonance-force
microscopy \cite{bib:StipePRL2001}, Raman transitions
\cite{bib:KoilerPRL2003}, far-infrared induced spin-to-charge
conversion \cite{bib:Hollenberg2003}, optical readout \cite{bib:JelezkoPRL2004} and the use of asymmetric confining potentials \cite{bib:FriesenPRL2004}.

The original Kane proposal for spin to charge conversion requires
two phosphorus donors: the qubit and a reference.  First, the
nuclear spin information is transferred to the electron spins. Then
spin-dependent tunneling between the qubit and reference is used to
determine whether the two spins are aligned parallel or
anti-parallel.  The tunneling creates a D$^+$D$^-$ system, where
the D$^+$ state is an ionized donor, and the D$^-$ state a doubly
occupied donor.  The change in the charge distribution between the
neutral and the D$^+$D$^-$ system is monitored with a single
electron transistor (SET) \cite{bib:GrabertNATO1992}. Although the
D$^+$D$^-$ state has been observed via far-infrared transmission
\cite{bib:CapizziSSC1979}, under the conditions required to
adiabatically form the D$^+$D$^-$ system in a top-gate controlled
structure, it appears that the state will be quasi-bound, with a
lifetime incompatible with SET readout \cite{bib:Hollenberg2003}. It
is therefore essential to determine alternative readout methods that
avoid the $D^-$ state problem.  We present such an alternative here.

\begin{figure}[tb]
\includegraphics[height = 5cm,clip]{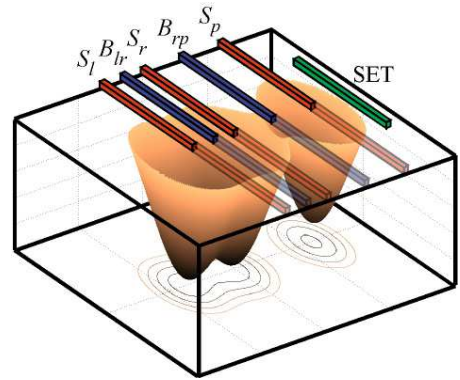}
\caption{\label{fig:Schematic} Schematic showing triple well
potential with top gates and readout SET.  The two leftmost wells
are strongly coupled, providing for a significant exchange
interaction, the third well is further removed so as to act as a
weakly coupled probe and is ionized prior to readout.  The
electron in the leftmost well is the spin qubit, $l$, the electron
in the central well is the reference spin, $r$, and the third well
is the probe site, $p$.}
\end{figure}

Our method is an all-electrical spin-to-charge conversion where an
extra, unoccupied site (the probe site) is introduced to
facilitate the readout.  The arrangement is illustrated in Fig.
\ref{fig:Schematic}.  There are three potential wells, which could
be derived from three donors, labelled $l$, $r$, and $p$, and two
electrons. The electron in $l$ is the qubit, the electron in $r$
is the reference spin, and $p$ is the probe site. The energies of
the states are controlled with shift gates, $S_{\alpha}$ ($\alpha
= l,r,p$) and the tunneling and exchange interactions between
sites are controlled with barrier gates $B_{lr}$ and $B_{rp}$. The
$lr$ system is strongly coupled and the $rp$ system weakly
coupled, so as to probe the $lr$ dynamics.

Readout is via a SET monitoring the population of the probe site.
Our scheme discriminates between singlet and triplet states of the
$lr$ system which is equivalent to measuring the spin of $l$ with
known $r$ \cite{bib:Kane}. We use the energy difference between the
singlet and triplet states of the combined qubit and reference
system to effect coherent tunneling into the probe, and hence qubit
read out.

We illustrate our method for the Kane quantum computer, but the scheme is
completely general and is applicable to most solid-state spin
quantum computing schemes (including those of Refs.
\cite{bib:LossPRA1998}, \cite{bib:VrijenPRA2000} and \cite{bib:SkinnerPRL2003}). More
generally, this scheme provides an alternative tool for examining
spin properties in quantum structures.  Ionicioiu and
Popescu \cite{bib:Ionicioiu} have proposed a different scheme using an
ancilla charge state to read out a spin qubit.

Earlier, we presented a scheme for readout of a charge-based QC
using a probe site \cite{bib:Greentree2003} (three-site,
one-electron case). Three-site two-electron models have been
proposed for entangled current formation
\cite{bib:EntangledCurrent}.   Such schemes differ
qualitatively from that discussed here.

Given the added complexity of fabricating a triple-donor system over
a more conventional two-donor system, and the increased coherence
times that will be required, it is wise to identify regimes where
our scheme is advantageous.  As mentioned above, if the lifetime of
the D$^+$D$^-$ system is less than the SET readout time, two-donor
spin to charge conversion will not be practical. Our scheme,
combined with the charge shelving described below, circumvents such
lifetime issues by avoiding the D$^-$ state. Note that for spins in
quantum dots \cite{bib:LossPRA1998} there are no problems due to
double occupancy of the states, but there may be utility in the
present proposal due to the lower required transfer potentials. We
also require the following inequalities to be satisfied, $J/\hbar
\gg \Omega_{rp} \sim 1/T_2^{rp}$ where $J$ is the exchange
interaction strength between electrons on sites $l$ and $r$ where we
have assumed $J \gg J_{rp}, J_{lp}$, $\Omega_{ij}$ is the coherent
tunneling rate on the $ij$ transition with $\Omega_{lp} = 0$, and
$1/T_2^{rp}$ is the (electrostatic) dephasing rate of charge motion
on the $rp$ transition.  This inequality will become clearer below,
however these inequalities are compatible with current thinking on
lifetimes of spin and charge qubits.

We assume the SET functions as a weak measurement device,
described by an effective $T_2$ time acting on the basis states of
electron occupation of the probe donor.  A full quantum treatment SET readout of a charge qubit has been
performed by Wiseman \textit{et al.} \cite{bib:WisemanPRB2001}.  We further assume that
this $T_2$ time is slow compared with the other timescales of the
system, except for the $T_1$ time and is therefore not treated in
our discussions.  These assumptions are physically realistic for
Si:P spin qubits, but more detailed calculations must be performed
to quantitatively determine the dynamics.

Detailed analyses of the effects of top gates on the spin states
of coupled two-site, two electron systems have been performed
\cite{bib:CoupledDot,bib:WellardPrePrint2003}.  These treatments
perform calculations to derive couplings in realistic systems.  We
make no attempt to replicate these important results, rather we
\emph{assume} the existence of appropriate interactions to illustrate the concepts of our scheme.

To understand the mechanism for transfer, we analyze the Hamiltonian
for the two-electron, three-site problem on the basis of states
$|\alpha \beta \bullet\rangle$, $|\alpha \bullet \beta \rangle$,
$|\bullet \alpha \beta \rangle$ for $\alpha, \beta = \downarrow,
\uparrow$ where the ordering is $l,r,p$ and $\bullet$ denotes an
unoccupied site.
\begin{eqnarray}
{\cal H} &=& \sum_{\alpha = \downarrow, \uparrow}
    \left[
        \sum_{i=l,r,p} E_i b^{\dag}_{i,\alpha} b_{i,\alpha} \right. \nonumber \\
        &+& \left. \hbar \Omega_{lr} \left( b^{\dag}_{l,\alpha} b_{r,\alpha} + h.c. \right)
        +  \hbar \Omega_{rp} \left( b^{\dag}_{r,\alpha} b_{p,\alpha} + h.c. \right) \right]  \nonumber \\
        &+& 4 J \sum_{i,j=l,r} S_i \cdot S_j    + g \mu_B B \sum_{i=l,r,p} S_i^z,
\end{eqnarray}
where $b_{i\alpha}$ is the annihilation operator for an electron on
site $i$ with spin $\alpha$; $B$ is the magnetic field; $E_i$ is the
electrostatic energy of an electron on site $i$; $S_i = b^{\dag}_{i
\downarrow} b_{i \uparrow}$; $S_i^z = (1/2) \left(b^{\dag}_{i
\uparrow} b_{i \uparrow} - b^{\dag}_{i \downarrow} b_{i
\downarrow}\right)$ and we define $B^* = g \mu_B B$ as the Zeeman
energy splitting.

\begin{figure}[tb]
\includegraphics[height = 10cm,clip]{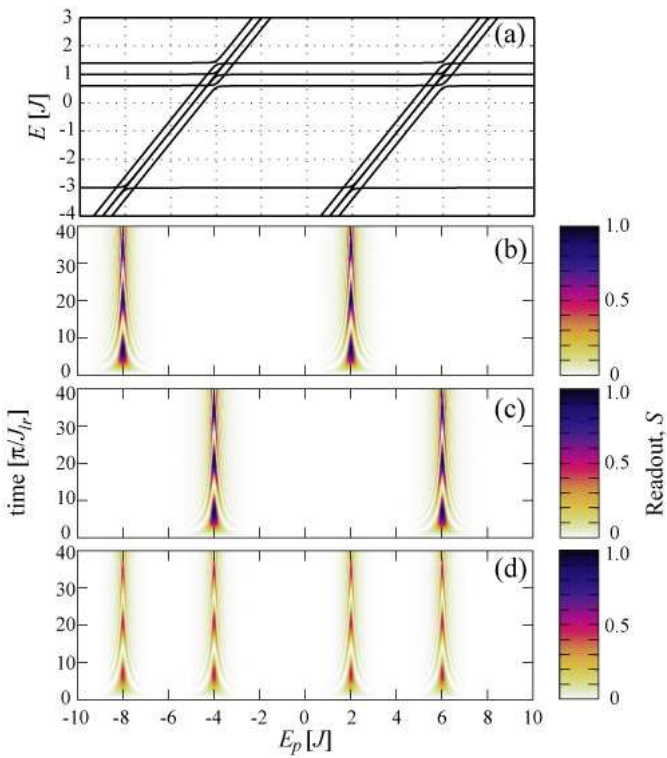}
\caption{\label{fig:Evals} (a) Eigenvalues for the
two-spin, three-well case as a function of the energy of the probe
state, $E_p$, for constant Zeeman splitting and exchange
interaction. The states with occupancy in the third site migrate
upwards in the figure.  Anti-crossings at biases corresponding to
resonance with the singlet and triplet states signify charge
transfer to the third well. All triplet anti-crossings appear at
the same $E_p$, indicating one cannot resolve the individual
triplet states. (b) - (d) Transient bias
spectroscopy showing readout variable ${\cal S}$ as a function of
$E_c/J$ and time (in units of $\pi\hbar/J$ for $B^*/J=0.4$,
$\Omega_{rp}/J = 0.1$, $E_l = E_r = 0$ and various initial
conditions, (b) any of the triplet states, (c) singlet state, and
(d) $\rho(0) = |\uparrow\downarrow\bullet\rangle$, which is a
superposition of singlet and one of the triplet states.}
\end{figure}

The eigenvalues of the system as a function of $E_p$ are illustrated
in Fig. \ref{fig:Evals} (a) where we have chosen $B^*=J/5$ and
$\Omega_{lr} = 5J$, so the singlet-triplet sub-manifolds are well
resolved and the symmetric and anti-symmetric manifolds with $p$
occupied [$(|\alpha \bullet \beta\rangle \pm |\bullet \alpha
\beta\rangle)/\sqrt{2}$, $\alpha,\beta = \uparrow,\downarrow$]
similarly well resolved. Anti-crossings in the evolution of the
eigenvalues indicate where the states change their character.  If
the system is initially prepared with electrons in sites $l$ and
$r$, then an anti-crossing for adiabatically swept $E_p$ corresponds
to electronic transfer from $r$ to $p$. There are two sets of
diagonal lines, corresponding to the final state of the electron in
the $lr$ system being in either the symmetric (lower energies) or
anti-symmetric (higher energy) superpositions.  Within each set,
there are two biases where charge transfer anti-crossings occur,
which correspond to transfer of an electron from the singlet state
at $E_p = \pm \Omega_{lr} -3J$, and from the triplet states at $E_p
= \pm \Omega_{lr} J$.  Note that all three triplet states within the
same charge symmetry sub-manifold anti-cross at the same bias,
suggesting that this scheme is not able to resolve the individual
triplet components.

To determine the expected readout, we take the trace of the density matrix,
$\rho$, over all states with a population in the probe dot $p$, which we
term $\mathcal{S}$.  We perform a transient
analysis of ${\cal S}$ by solving the density matrix equations of
motion, $\dot{\rho} = (-i/\hbar) [{\cal H},\rho]$, for various
initial conditions as a function of probe bias.

Results showing ${\cal S}$ as a function of bias and time are
presented in Fig. \ref{fig:Evals} (b) - (d).  The figures show
coherent oscillations in ${\cal S}$, peaking at the resonance bias,
with the oscillation frequency increasing away from resonance.  Fig.
\ref{fig:Evals} (b) shows the bias spectroscopy when the system is
initialized in the singlet state,
$(1/\sqrt{2})(|\uparrow\downarrow\bullet\rangle -
|\downarrow\uparrow\bullet\rangle)$.  Population is transferred to
the probe when $E_p = \pm \Omega_{lr} - 3J$. Fig. \ref{fig:Evals}
(c) shows $\mathcal{S}$ for the system initialized in either of the
triplet states, $|\uparrow\uparrow\bullet\rangle$,
$(1/\sqrt{2})(|\uparrow\downarrow\bullet\rangle +
|\downarrow\uparrow\bullet\rangle)$ and
$|\downarrow\downarrow\bullet\rangle$. This state has a different
spectroscopic signature, with readout observed at $E_p = \pm
\Omega_{lr} + J$, with the individual triplet states are
unresolvable.  In Fig. \ref{fig:Evals} (d) we present results
obtained when the initial state was
$|\uparrow\downarrow\bullet\rangle$. This state corresponds to a
superposition of the singlet state and the symmetric state, i.e.
$|\uparrow\downarrow\bullet\rangle = (1/\sqrt{2})
\left[\left(|\uparrow\downarrow\bullet\rangle +
|\downarrow\uparrow\bullet\rangle \right) +
\left(|\uparrow\downarrow\bullet\rangle -
|\downarrow\uparrow\bullet\rangle \right)\right]$. Because of this
superposition, we observe two sets of biases (one per submanifold)
where charge transfer to the probe donor is observed, and this is
clearly seen in the spectroscopic signature presented in Fig.
\ref{fig:Evals} (d).  Unlike the charge-qubit readout
\cite{bib:Greentree2003}  and optical case
\cite{bib:GreentreePRA2002}, there is no interference between the
features in Fig. \ref{fig:Evals} (d).  This is because there are no
shared final states in the spin readout scheme, and therefore no
interference.  Inclusion of a measurement induced $T_2$ here will
tend to wash out the oscillations in all cases shown in Fig.
\ref{fig:Evals}, allowing $\mathcal{S}$ to evolve to a steady state
in a time commensurate with $T_2$.

For single-shot readout for a QC, we propose implementing a form of
adiabatic fast passage (AFP) \cite{bib:VitanovARPC2001} on the $rp$
transition.  This is analogous to an earlier suggestion for
single-shot readout of a charge-qubit in the superposition basis
\cite{bib:Greentree2003}. The advantages of AFP over bias
spectroscopy include insensitivity to coherent oscillations on the
$r-p$ transition and robustness to gate errors.  Although we do not
discuss decoherence in this work, it is important to realize that
the minimum length of time to implement an AFP gate sweep will be of
order $10 \pi\hbar /J$. Thus the decoherence time should be long
compared to this timescale. Given the already demanding requirements
for dephasing in QCs\cite{bib:PreskillPRSLA1998}, i.e. that the
decoherence time should be $10^3-10^6$ times the coherent
oscillation time, $\hbar/J$, then if we \emph{assume} that
construction of a scalable qubit is possible, the added overhead of
implementing the AFP sweep is negligible.

\begin{figure}[tb]
\includegraphics[height = 4.5cm,clip]{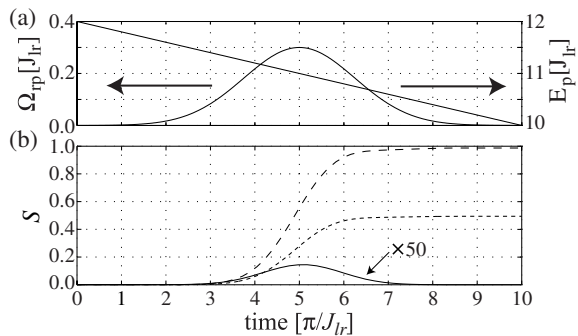}
\caption{\label{fig:AFP} Adiabatic fast passage (AFP) trajectories
and readout for various initial conditions. (a) shows
$\Omega_{rp}$ (left axis) and $E_p$ (right axis) as a function of time for the
AFP sequence.  (b) shows ${\cal S}$ as a function of time for the
$l-r$ system in the singlet state (solid line), triplet state
(long dashes) and a superposition of singlet and triplet states
(short dashes).  The singlet readout has been multiplied by $50$
to be visible on this scale. ${\cal S}$ migrates smoothly to the
required value suggesting this is an appropriate mechanism for
performing readout.}
\end{figure}

To effect the AFP gate sweep, we vary $E_p$ and $\Omega_{rp}$ according to
\begin{eqnarray}
   E_p &=& \Omega_{lr} + 2 J_{lr} \left(1-t/t_{\max}\right), \nonumber\\
    \Omega_{rp} &=& \Omega_{rp}^{\max}
    \left[ 1 - \cos\left(2\pi t / t_{\max} \right) \right]/2,
\end{eqnarray}
where $\Omega_{rp}^{\max} = 0.3 J_{lr}/\hbar$, $t_{\min}=0$, $\Omega_{lr} = 10 J_{lr} /\hbar$ and
$t_{\max} = 10 \pi \hbar/ J_{lr}$. To illustrate this, Fig.
\ref{fig:AFP} (a) shows $\Omega_{rp}(t)$ (left axis) and $E_p$ (right axis).  Note that in keeping
with conventional AFP schemes, the scheme is fairly insensitive to
the exact form of $\Omega_{rp}$.  ${\cal S}$ as a function of time is presented in \ref{fig:AFP} (b)
for the three cases of the $l-r$ system being initially in the
singlet state (solid line), triplet state (long dashes) and
superposition state $|\uparrow\downarrow\bullet\rangle$ (short
dashes).

With the addition of gate noise and decoherence, nonadiabatic
techniques would only be expected to transfer on average half an
electron to the probe site.  By contrast, the AFP scheme will
transfer a full electron (to arbitrary precision) even in the
presence of gate errors.  Therefore this scheme is compatible with
single-shot readout, whereas the nonadiabatic scheme is not,
yielding only a statistical result.

Until now we have deliberately concentrated on an arbitrary system
to highlight the generality of our readout mechanism.  We conclude
by turning our attention specifically to readout of a Kane-type QC
\cite{bib:Kane}.  For $lr$ donor spacing of $\sim 15\mathrm{nm}$,
the potential on a B gate required to shift the exchange coupling
from $J=0$ to $J\sim 0.1 \mathrm{meV}$ is $1\mathrm{V}$
\cite{bib:WellardPrePrint2003}.  At these separations, we would
expect $\Omega_{lr} \sim 1\mathrm{meV}$.  To achieve the bias sweep
necessary our AFP protocol, we would need to vary $E_p$ smoothly
from $E_p = 1.2 \mathrm{meV}$ to $E_p = 1.0 \mathrm{meV}$. TCAD
Modelling \cite{bib:TCAD} suggests that an S gate to the right of
the donor for a charge qubit will shift the potential by $\sim
4\mathrm{meV}$ for a change in gate potential of $1\mathrm{V}$. This
implies that the S gate potential must be controlled to of order
tens of mV, which is achievable using conventional technology.  The
requirement for $\Omega^{\max}_{rp} \sim 0.3 J_{lr}$ would imply a
spacing between $r$ and $p$ of around $25 \mathrm{nm}$, again
achievable with current technology. Other QC schemes will have quite
different site-gate couplings due to different geometries.  One
would normally expect larger couplings for schemes where the quantum
sites are extended structures (e.g. GaAs quantum dots, where quantum
coherence has been shown \cite{bib:OosterkampNature1998}) than for
the single donors envisaged here.

In summary, we have presented a high-fidelity, single-shot scheme
for performing readout of a spin qubit, making use of a reference
qubit and an empty probe site.  The energy difference between the
singlet and triplet states of the spin-reference system is probed
using bias spectroscopy to the probe site, and the change in charge
on the probe is monitored with a SET.  This constitutes a form of
spin-to-charge conversion and charge shelving, where the spin
information is transferred to the charge of a long-lived probe site.
This enables the use of a measurement device where the measurement
time is longer than the coherence time of the qubit.  Our techniques
should be applicable to a wide range of different spin-based quantum
computing schemes.

The authors would like to thank G.~Milburn, F.~Green and C.~J.~Wellard for
useful discussions. This work was supported by the Australian
Research Council, the Australian government and by the US National
Security Agency (NSA), Advanced Research and Development Activity
(ARDA) and the Army Research Office (ARO) under contract number
DAAD19-01-1-0653.


\end{document}